\documentstyle[aps,preprint]{revtex}
\begin{document}
\title{On the string-inspired approach to QED in external field}
\author{R.~ SHAISULTANOV\thanks{%
Email:shaisultanov@inp.nsk.su}}
\address{Budker Institute of Nuclear Physics \\
630090, Novosibirsk 90, Russia}
\maketitle

\begin{abstract}
Strassler's formulation of the string-derived Bern-Kosower formalism is
extended to consider QED processes in homogeneous constant external
field. A compact expression for the contribution of the one-loop with
arbitrary number of external photon lines is given for scalar QED.
Extension to spinor QED is shortly discussed.
\end{abstract}

\newpage
\narrowtext

Four years ago Bern and Kosower \cite{h} used string theory to derive a new
formalism for the calculation of one-loop amplitudes in field theory.
Inspired by this work Strassler derived a set of rules for computing
one-loop Green functions in field theory directly from one-dimensional path
integrals for relativistic point-particle models\cite{as} . Earlier similar
results were obtained by Polyakov \cite{ls}. The method has been used and
extended by several authors so as to allow for example the calculation of
effective actions \cite{k} and of higher loops \cite{mv} . Also Lam \cite{lm}
showed that expressions similar to Bern-Kosower rules can be obtained
using the Feynman-parameter representation even for multiloop
diagrams in QED.

Here we wish to extend this method to study QED processes in external
constant electromagnetic fields. We will consider processes described by
one-loop diagrams with N external photons. For example the N=2 case will
give us the polarization operator \cite{v,br1}, N=3 case describe
photon splitting in external electromagnetic field \cite{adler}.
Since nonperturbative calculation in $F_{\mu \nu }$ has to be
employed , the calculations with external fields are highly
nontrivial. For the comprehensive presentation of QED with external
fields and the technique used see for example \cite{tech}.

We will begin with modified Strassler's expression for amplitude with
external field $A_\mu (x)$ added
\begin{equation}
\label{eq:shock}
\begin{array}{c}
\Gamma _N\left( k_1,...,k_N\right) =-(i\;g)^N\;\int\limits_0^\infty \;
\frac{dT}T\;{\cal N\;}\int {\cal D}x\exp \left\{ -\int d\tau \left[ \frac{%
\dot x^2}{2{\cal E}}-igA(x)\cdot \dot x-m^2\right] \right\} \times \\ \times
\prod_{i=1}^N\int\limits_0^Tdt_i\varepsilon _i\cdot \dot x%
(t_i)\;e^{i\;k_i\cdot x(t_i)}
\end{array}
\end{equation}

To calculate this expectation value we will use standard path integral
methods of string perturbation theory \cite{witten}.We will get
\begin{equation}
\label{eq:fas}
\begin{array}{c}
\Gamma _N\left( k_1,...,k_N\right) =-(i\;g)^N\;\int\limits_0^\infty \;
\frac{dT}T\prod_{i=1}^N\int\limits_0^Tdt_i\;{\cal N\;}\int {\cal D}x\exp
\left\{ -\int\limits_0^Td\tau \left[ \frac{\dot x^2}{2{\cal E}}-igA(x)\cdot
\dot x-m^2\right] \right\} \times  \\ \exp \left. \left\{
+\int\limits_0^Td\tau \;J_\mu (\tau )x^\mu \left( \tau \right) \right\}
\right| _{linear\;in\;each\;\varepsilon }
\end{array}
\end{equation}

where the source for x$^\mu $ is
\begin{equation}
\label{eq:gee}J^\mu (\tau )=\sum_{i=1}^N\delta \left( \tau -t_i\right)
\left( \varepsilon _i^\mu \frac \partial {\partial t_i}+i\;k_i^\mu \right)
\end{equation}

We take $A_\mu \left( x\right) =-\frac 12F_{\mu \nu }x^\nu $ then
integration over $x\left( \tau \right) $ gives
\begin{equation}
\label{eq:di}
\begin{array}{c}
\Gamma _N\left( k_1,...,k_N\right) =-(i\;g)^N\;\int\limits_0^\infty \;
\frac{dT}T\prod_{i=1}^N\int\limits_0^Tdt_i[2\pi {\cal E}T]^{-2}\frac{\left| -%
\frac 14\partial ^2\right| ^{\frac 12}}{\left| -\frac 14\partial ^2-\frac 12%
igF\otimes \partial \right| ^{\frac 12}}\times \\ \exp \left\{ -\frac 12%
\int\limits_0^Td\tau \int\limits_0^Td\tau ^{\prime }J^\mu (\tau )G_{\mu \nu
}\left( \tau ,\tau ^{\prime }\right) J^\nu (\tau ^{\prime })+\frac{{\cal E}}2%
m^2T\right\}
\end{array}
\end{equation}

The matrix Green function satisfies the equation
\begin{equation}
\label{eq:cov}\frac 1{{\cal E}}\ddot G+igF\dot G=\delta \left( \tau -\tau
^{\prime }\right) -\frac 1T
\end{equation}
which has the solution when the condition of periodicity in $\tau
\rightarrow \tau +T$ is imposed
\begin{equation}
\label{eq:gw}G_{\mu \nu }\left( \tau ,\tau ^{\prime }\right) =\left.
\begin{array}{c}
-i
\frac{e^{-ig{\cal E}F(\tau -\tau ^{\prime })}-1}{gF\left( e^{-ig{\cal E}%
FT}-1\right) }+\frac i{gFT}(\tau -\tau ^{\prime
})\;\;\;\;\;\;\;\;\;\;\;\;\;at\;\;\;\;\;\tau >\tau ^{\prime } \\ -i\frac{%
e^{-ig{\cal E}F(T+\tau -\tau ^{\prime })}-1}{gF\left( e^{-ig{\cal E}%
FT}-1\right) }+\frac i{gFT}(\tau -\tau ^{\prime })\;+\frac i{gF}%
\;\;\;at\;\;\;\;\;\tau <\tau ^{\prime }
\end{array}
\right. \;
\end{equation}

In limit F=0 it give the old Green function of \cite{as}. Using now for
determinants results of \cite{k} and choosing the gauge ${\cal E}=2$ we
obtain

\begin{equation}
\label{eq:sec}
\begin{array}{c}
\Gamma _N\left( k_1,...,k_N\right) =-(i\;g)^N\;\int\limits_0^\infty \;
\frac{dT}{(4\pi )^2T}\frac{g^2a\;b}{\sin (igbT)\sinh (igaT)}%
\prod_{i=1}^N\int\limits_0^Tdt_i\times \\ \exp \left\{ -\frac 12%
\int\limits_0^Td\tau \int\limits_0^Td\tau ^{\prime }J^\mu (\tau )G_{\mu \nu
}\left( \tau ,\tau ^{\prime }\right) J^\nu (\tau ^{\prime })+m^2T\right\}
\end{array}
\end{equation}

where
\begin{equation}
\label{eq:vier}
\begin{array}{c}
a^2=(
{\cal F}^2+{\cal J}^2)^{\frac 12}+{\cal F}\;\;,b^2=({\cal F}^2+{\cal J}^2)^{%
\frac 12}-{\cal F} \\ {\cal F}=-\frac 14F_{\mu \nu }F^{\mu \nu },\;\;{\cal J}%
=-\frac 14F_{\mu \nu }^{*}F^{\mu \nu }
\end{array}
\end{equation}
Now we have as a final result
\begin{equation}
\label{eq:a}
\begin{array}{c}
\Gamma _N\left( k_1,...,k_N\right) =-(i\;g)^N\;\int\limits_0^\infty \;
\frac{dT}{(4\pi )^2T}\frac{g^2a\;b}{\sin (igbT)\sinh (igaT)}%
e^{m^2T}\prod_{i=1}^N\int\limits_0^Tdt_i\times \\ \left. \exp \frac 12%
\sum_{i,j}^N\left\{ k_iG(t_i-t_j)k_j-i\varepsilon _i\frac \partial {\partial
t_i}G(t_i-t_j)k_j-ik_i\frac \partial {\partial t_j}G(t_i-t_j)\varepsilon
_j-\varepsilon _i\frac{\partial ^2}{\partial t_i\partial t_j}%
G(t_i-t_j)\varepsilon _j\right\} \right| _{linear\;in\;each\;\varepsilon }
\end{array}
\end{equation}
where due to translation invariance the integration over N +1 variables is
really over N variables.Here for symmetry we does not remove this redundant
integration though this can be made at any step of calculation. In the case
N=2 eq.(\ref{eq:a}) reproduce results of \cite{br1}.

Extension to spinor case is straightforward : we need only new additional
Green function G$_F$ \cite{as} which now obey

\begin{equation}
\label{eq:b}\dot G_F+ig{\cal E}FG_F=2\delta (\tau -\tau ^{\prime })
\end{equation}

which has a solution when the condition of antiperiodicity in $\tau
\rightarrow \tau +T$ is imposed
\begin{equation}
\label{eq:fer}G_F\left( \tau ,\tau ^{\prime }\right) =\left.
\begin{array}{c}
\;\;\;\;\;\;\;2
\frac{e^{-ig{\cal E}F(\tau -\tau ^{\prime })}}{e^{-ig{\cal E}FT}+1}%
\;\;\;\;\;\;\;\;\;\;\;at\;\;\;\;\;\tau >\tau ^{\prime } \\ 2\frac{e^{-ig%
{\cal E}F(T+\tau -\tau ^{\prime })}}{e^{-ig{\cal E}FT}-1}\;\;\;\;\;\;\;\;\;%
\;\;at\;\;\;\;\;\tau <\tau ^{\prime }
\end{array}
\right.
\end{equation}

Note finally that calculations with functions of the matrix $F$ can
be easily done using techniques presented in the appendix of paper
\cite{br2}.

\end{document}